\documentclass{iopconfser}
\usepackage{amsmath} 
\usepackage{graphicx}
\usepackage{xcolor}
\usepackage{orcidlink}
\usepackage{etoolbox}
\newtoggle{checklength}
\toggletrue{checklength} 
\usepackage[normalem]{ulem}
\usepackage{hyperref}
\usepackage{doi}

\begin{document}
\title{
Spin-induced quadrupole moment test for eccentric compact binaries
}

\author{
  Syed~U.~Naqvi$^{1,2}$\orcidlink{0000-0002-9380-0773},
  Chandra~Kant~Mishra$^{1,2}$\orcidlink{0000-0002-8115-8728}
}

\iftoggle{checklength}{
  \affil{$^1$Department of Physics, Indian Institute of Technology Madras, Chennai 600036, India}

  \affil{$^2$
  Centre for Strings, Gravitation and Cosmology, Department of Physics, Indian Institute of Technology Madras, Chennai 600036, India}

  \email{umair.kmc@gmail.com, ckm@physics.iitm.ac.in}
}{}

\begin{abstract}
Spin-induced deformations of individual components of a binary can be quantified using the gravitational wave signal the binary emits. Such deformations are characterised by a parameter, $\kappa$, which takes a value of 1 for a black hole and thus its measurement can be used to test the black hole nature. However, in practice, only a symmetric combination of this parameter for a binary ($\kappa_s$) can be measured, thus enabling a test for the black hole nature as studied previously in literature. While previous studies have focused on circular binaries, our work extends the proposal to test the black hole nature of compact objects to eccentric systems, and we forecast the expected measurement precision using a Fisher matrix analysis. We find that, the error in the measurement of the parameter $\kappa_s$ reduces from a value of $\sim$18\% (for the circular case) to close to 8\%($4$\%) for a $10M_{\odot}$ system with dimensionless component spins $>0.8$ and with a reference initial eccentricity ($e_0$) of 0.2(0.5) evaluated at 5Hz for Cosmic Explorer(CE). Compared to the estimates obtained by using advanced LIGO design sensitivity, eccentricity and overall improved sensitivity of CE detectors together seem to improve these estimates almost by an order of magnitude.
\end{abstract}
\vspace{-25pt}
\section{Introduction}
In general relativity, the spin-induced quadrupole moment (SIQM) of a compact object provides an important avenue to probe its nature, offering a way to distinguish black holes from other exotic compact objects. For an object of mass $m_A$ and dimensionless spin $\chi_A$, the SIQM can be expressed as $Q_{A} = -\kappa \chi_A^{2} m_A^{3}$, where the parameter $\kappa$ characterizes the quadrupolar deformation for the object and the label $A=1,2$ corresponding to the two constituents. Within the Kerr solution, the no-hair theorem fixes $\kappa = 1$, while for neutron stars $\kappa$ typically falls in the range $2\text{--}14$~\cite{Laarakkers:1997hb, Pappas:2012ns}, for boson stars in range $\sim$10-100~\cite{Ryan:1996nk, Liebling:2012fv} and can take -ve values for gravastars~{\cite{Mazur:2004fk,Yang:2022gic}.
A measurement of $\kappa$ thus serves as a null test of the black hole hypothesis. Previous works have investigated this possibility in the context of binaries on circular orbits~\cite{Krishnendu:2017shb, Krishnendu:2019ebd, Divyajyoti:2021uty, LIGOScientific:2019fpa}. In this study, we build on these efforts by extending the test to eccentric binaries in the context of a third generation detector -- Cosmic Explorer~\cite{2021arXiv210909882E}.
\vspace{-15pt}
\section{Waveforms}
We employ a frequency domain, inspiral model, based on stationary phase approximation (SPA), of ~\cite{Sridhar:2024zms} providing 3PN accurate phasing expressions for spinning compact binaries on eccentric orbits. Reference~\cite{Sridhar:2024zms} extended by including the spin information to an earlier work~\cite{Moore:2016qxz} which assumed nonspinning components. We further augment this with 2PN accurate prescription for the dominant harmonic (quadrupolar) amplitude and orbital phase through 4PN order for the circular part~\cite{Krishnendu:2017shb, Mishra:2016whh}.
Note also, the eccentricity related corrections of \cite{Sridhar:2024zms} are accurate to $\mathcal{O}(e_0^8)$ which means only terms beyond the 8th power in eccentricity are neglected. It was shown in Ref.~\cite{Sridhar:2024zms} that, such a model should be accurate enough to analyze systems with eccentricities up to $\sim$0.5. The structure of the waveform employed takes the following form 
\begin{equation}
\tilde{h}_{\rm SPA}(f) = \frac{M^2}{D_L} \sqrt{\frac{5\pi\eta}{48}} 
\sum_{n=0}^{4}\sum_{k=0}^6 V_k^{\,n-7/2} C_{k}^{(n)} 
e^{i\left[k\Psi(f/k) - \pi/4\right]},
\end{equation}
In the above, $M$ is the total detector-frame mass, $\eta$ is the symmetric mass ratio parameter, and 
$D_L$ is the luminosity distance to the source. 
The quantities $V_k$ and $C_k^{(n)}$ denote post-Newtonian coefficients associated with the $k^{\rm th}$ harmonic of the waveform, while $\Psi(f/k)$ represents the orbital phase evolution. 
In this work, we restrict ourselves only to the dominant harmonic ($k=2$). 
The spin-induced quadrupole parameters ($\kappa$)
enter both the phase 
and the amplitude; see Refs.~\cite{Krishnendu:2017shb,Mishra:2016whh} for explicit expressions. 

\vspace{-12pt}
\begin{figure}[h!]
    \centering
    \includegraphics[width=0.5\textwidth]{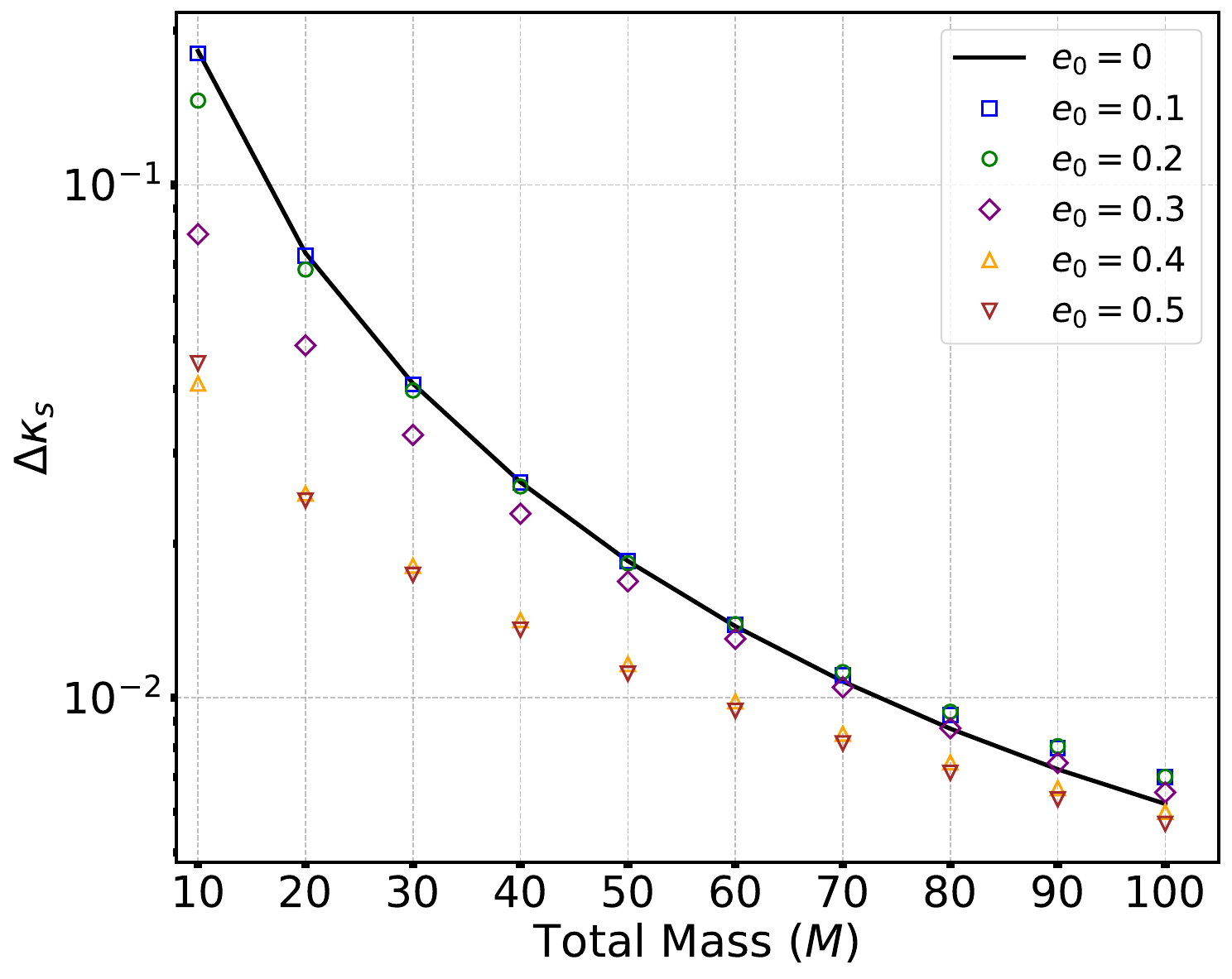}
    \caption{Statistical $1-\sigma$ (one standard deviation) uncertainties on the SIQM parameter ($\kappa_s$) as a function of the total detector-frame mass $M$ for set of reference eccentricity values at 5Hz. The component mass ratio ($q=m_1/m_2$) is fixed to a value of 1.25 and component spins are chosen to have a value of 0.9 and 0.8 for heavier and lighter component while the binary's luminosity distance is fixed at 400 Mpc. Systems are analyzed using a Cosmic Explorer noise power spectral density (PSD)~\cite{Srivastava:2022slt}.
    }
    \label{fig:kappa_plot}
\end{figure}
\vspace{-28pt}
\section{Analysis and Results}
\label{sec:fisher}
We use a Fisher information matrix approach~\cite{Cutler:1994ys} to forecast the statistical uncertainties in the parameter measurements.\footnote{See Ref.~\cite{Vallisneri:2007ev} for possible caveats of the method.} The parameter space explored in this work is $\{\mathcal{M}, \eta, \chi_{1}, \chi_{2}, t_c, \phi_c, e_0, \kappa_s\}$
where $\mathcal{M}$ is the chirp mass, $\eta$ the symmetric mass ratio, 
$\chi_{1,2}$ the dimensionless spin parameters of the binary components, 
$t_c$ and $\phi_c$ the time and phase at coalescence, 
$e_0$ orbital eccentricity at a reference frequency of 5 Hz (except in the circular case)\footnote{In this work $e_0$ refers to the eccentricity parameter associated with the time coordinate of the quasi-Keplerian (QK) approach~\cite{Moore:2016qxz}. Its evolution is governed by the PN equations in Eqs.~(19)–(20) of Ref.~\cite{Sridhar:2024zms}.}, and $\kappa_s$ the symmetric spin-induced quadrupole moment (SIQM) parameter.
The analysis is performed in context of the third generation detector -- Cosmic Explorer\cite{Reitze:2019iox} and for comparison also in context of Advanced LIGO~\cite{AdvancedLIGO2010}.      

Figure~\ref{fig:kappa_plot} displays our results for the measurements of the SIQM parameter. 
    The general trend is a \textit{decrease} in $\Delta{\kappa_s}$ with increasing total mass ($M$) and can be attributed to higher signal-to-noise ratio possible for heavier systems~\cite{Krishnendu:2019ebd}.\footnote{Errors should eventually increase when the SNR drops as we keep increasing the mass.}We note that this trend primarily reflects the scaling with SNR at a fixed distance of 400 Mpc.
    At the low mass end, including eccentricity improves the measurement of the SIQM parameter significantly. For instance, the error reduces from $\sim$18\% (for circular case) to almost 8\% for $e_0=0.2$ is again halved for $e_0=0.5$. Further, as expected, the over all reduced sensitivity of CE over advanced LIGO leads to significant improvements of the SIQM parameter. We find that, the error on $\kappa_s$ (again for the 10$M_{\odot}$ system) increase nearly 20-fold ($\sim500\%$) for the circular case with advanced LIGO sensitivity.
    
We looked into the impact of including eccentricity on the measurement of the SIQM parameter that characterizes spin-induced deformations of the binary constituents and thus offers a test of the black hole nature. We find that inclusion of eccentricity significantly improves the measurements compared to the circular case making a case of reanalysis of observed data for selected events with eccentric models. 

\section*{Acknowledgments} 
S.U.N. would like to thank  K. G. Arun and N. V. Krishnendu for useful
discussions. C.K.M. acknowledges the support of SERB’s Core Research Grant No.\ CRG/2022/007959. 
\bibliography{references}
\bibliographystyle{iopart-num}
\end{document}